\documentclass[runningheads]{llncs}
\usepackage{graphicx}
\usepackage{subcaption}
\usepackage{cite}
\usepackage{acro}
\usepackage{float}
\usepackage{xcolor}
\usepackage{amsmath}
\usepackage{multirow}
\usepackage{multicol}
\usepackage{wrapfig}
\usepackage{booktabs}   
\usepackage{subcaption}
\usepackage{wrapfig}
\usepackage{float}
\usepackage[english]{babel}
\usepackage{hyperref}
\usepackage{adjustbox}
\usepackage{amssymb}

\begin{document}
\title{Scaling Up 3D Kernels with Bayesian Frequency Re-parameterization for Medical Image Segmentation}
\titlerunning{RepUX-Net}
% \author{No Author Shown}
\author{Ho Hin Lee\inst{1}\and Quan Liu\inst{1}\and Shunxing Bao\inst{1} \and Qi Yang\inst{1} \and Xin Yu \and Leon Y. Cai\inst{3} \and Thomas Li\inst{3}\and Yuankai Huo\inst{1,2}\and Xenofon Koutsoukos\inst{1} \and Bennett A. Landman\inst{2}}
%
% \authorrunning{**** et al.}

% \institute{No Institute Shown}
% \newline
% \email{****@****.****}}
\institute{
Department of Computer Science, Vanderbilt University, Nashville, TN 37212, USA \and
Department of Electrical and Computer Engineering, Vanderbilt University, Nashville TN 37212, USA \and
Department of Biomedical Engineering, Vanderbilt University, Nashville TN 37212, USA}
\maketitle  % typeset the header of the contribution
\begin{abstract}
With the inspiration of vision transformers, the concept of depth-wise convolution revisits to provide a large Effective Receptive Field (ERF) using Large Kernel (LK) sizes for medical image segmentation. However, the segmentation performance might be saturated and even degraded as the kernel sizes scaled up (e.g., $21\times 21\times 21$) in a Convolutional Neural Network (CNN). We hypothesize that convolution with LK sizes is limited to maintain an optimal convergence for locality learning. While Structural Re-parameterization (SR) enhances the local convergence with small kernels in parallel, optimal small kernel branches may hinder the computational efficiency for training. In this work, we propose RepUX-Net, a pure CNN architecture with a simple large kernel block design, which competes favorably with current network state-of-the-art (SOTA) (e.g., 3D UX-Net, SwinUNETR) using 6 challenging public datasets. We derive an equivalency between kernel re-parameterization and the branch-wise variation in kernel convergence. Inspired by the spatial frequency in the human visual system, we extend to vary the kernel convergence into element-wise setting and model the spatial frequency as a Bayesian prior to re-parameterize convolutional weights during training. Specifically, a reciprocal function is leveraged to estimate a frequency-weighted value, which rescales the corresponding kernel element for stochastic gradient descent. From the experimental results, RepUX-Net consistently outperforms 3D SOTA benchmarks with internal validation (FLARE: 0.929 to 0.944), external validation (MSD: 0.901 to 0.932, KiTS: 0.815 to 0.847, LiTS: 0.933 to 0.949, TCIA: 0.736 to 0.779) and transfer learning (AMOS: 0.880 to 0.911) scenarios in Dice Score. Both codes and pretrained models are available at: \url{https://github.com/MASILab/RepUX-Net}

\end{abstract}
\section{Introduction}
With the introduction of Vision Transformers (ViTs), CNNs have been greatly challenged as seen with the leading performance in multiple volumetric data benchmarks, especially for medical image segmentation \cite{hatamizadeh2022unetr, hatamizadeh2022swin, tang2022self, zhou2021nnformer}. The key contribution of ViTs is largely credited to the large Effective Receptive Field (ERF) with a multi-head self-attention mechanism \cite{dosovitskiy2020image}. Note the attention mechanism is computationally unscalable with respect to the input resolutions \cite{liu2021swin, liu2022convnet}. Therefore, the concept of depth-wise convolution is revisited to provide a scalable and efficient feature computation with large ERF using large kernel sizes (e.g., $7\times 7\times 7$) \cite{liu2022convnet, lee20223d}. However, either from prior works or our experiments, the model performance becomes saturated or even degraded when the kernel size is scaled up in encoder blocks \cite{ding2022scaling, liu2022more}. We hypothesize that scaling up the kernel size in convolution may limit the optimal learning convergences across local to global scales. Recently, the feasibility of leveraging large kernel convolutions (e.g., $31\times 31$ \cite{ding2022scaling}, $51\times 51$\cite{liu2022more}) has been shown with natural image domain with Structural Re-parameterization (SR), which adapts Constant-Scale Linear Addition (CSLA) block (Fig. 2b) and re-parameterizes the large kernel weights during inference \cite{ding2022scaling}. As convolutions with small kernel sizes converge more easily, the convergence of small kernel regions enhances in the re-parameterized weight, as shown in Fig. 1a. With such observation, we further ask: \textbf{Can we adapt variable convergence across elements of the convolution kernel during training, instead of regional locality only?}

\begin{figure}[t!]
\centering
\includegraphics[width=\textwidth]{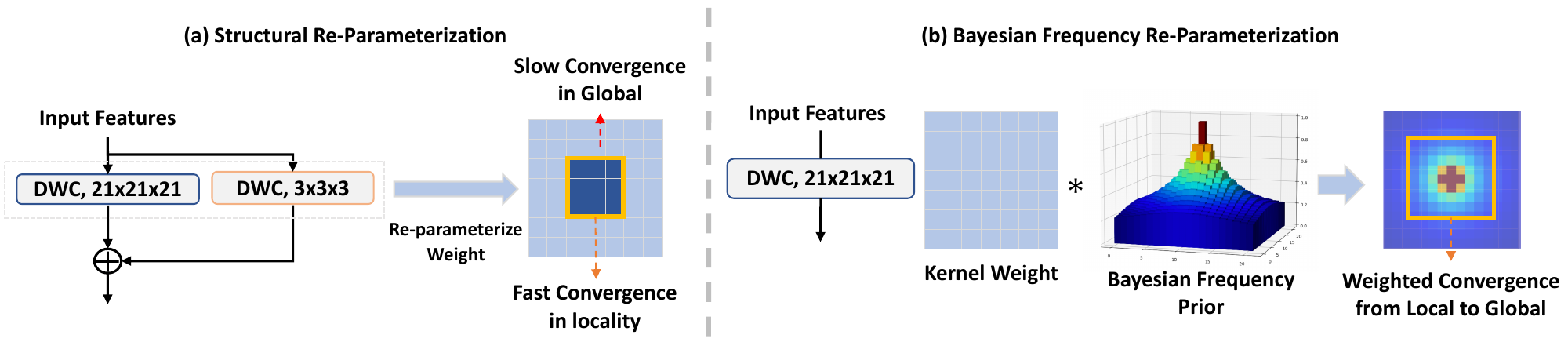}
\caption{With the fast convergence in small kernels, SR merges the branches weights and enhances the locality convergence with respect to the kernel size (deep blue region), while the global convergence is yet to be optimal (light blue region). By adapting BFR, the learning convergence can rescale in an element-wise setting and distribute the learning importance from local to global.} \label{idea_figure}
\end{figure}

In this work, we first derive and extend the theoretical equivalency of the weight optimization in the CSLA block. We observe that the kernel weight of each branch can be optimized with variable convergence using branch-specific learning rates. Furthermore, the ERF with SR is visualized to be more widely distributed from the center element to the global surroundings \cite{ding2022scaling}, demonstrating a similar behavior to the spatial frequency in the human visual system \cite{kulikowski1982theory}. Inspired by the reciprocal characteristics of spatial frequency, we model the spatial frequency as a Bayesian prior to adapt variable convergence of each kernel element with stochastic gradient descent (Fig. 1b). Specifically, we compute a scaling factor with respect to the distance from the kernel center and multiply the corresponding element for re-parameterization during training. Furthermore, we simplify the encoder block design into a plain convolution block only to minimize the computation burden in training and achieve State-Of-The-Art (SOTA) performance. We propose RepUX-Net, a pure 3D CNN with the large kernel size (e.g., $21\times 21\times 21$) in encoder blocks, to compete favorably with current SOTA segmentation networks. We evaluate RepUX-Net on supervised multi-organ segmentation with 6 different public volumetric datasets. RepUX-Net demonstrates significant improvement consistently across all datasets compared to all SOTA networks. We summarize our contributions as below:

\begin{itemize}
    \item We propose RepUX-Net with better adaptation in large kernel convolution than 3D UX-Net, achieving SOTA performance in 3D segmentation. To our best knowledge, this is the first network that effectively leverages large kernel convolution with plain design in the encoder for 3D segmentation.
    \item We propose a novel theory-inspired re-parameterization strategy to scale the element-wise learning convergence in large kernels with Bayesian prior knowledge. To our best knowledge, this is the first re-parameterization strategy to adapt 3D large kernels in the medical domain.
    \item We leverage six challenging public datasets to evaluate RepUX-Net in 1) direct training and 2) transfer learning scenarios with 3D multi-organ segmentation. RepUX-Net achieves significant improvement consistently in both scenarios across all SOTA networks.
\end{itemize}

\begin{figure}[t!]
\centering
\includegraphics[width=0.95\textwidth]{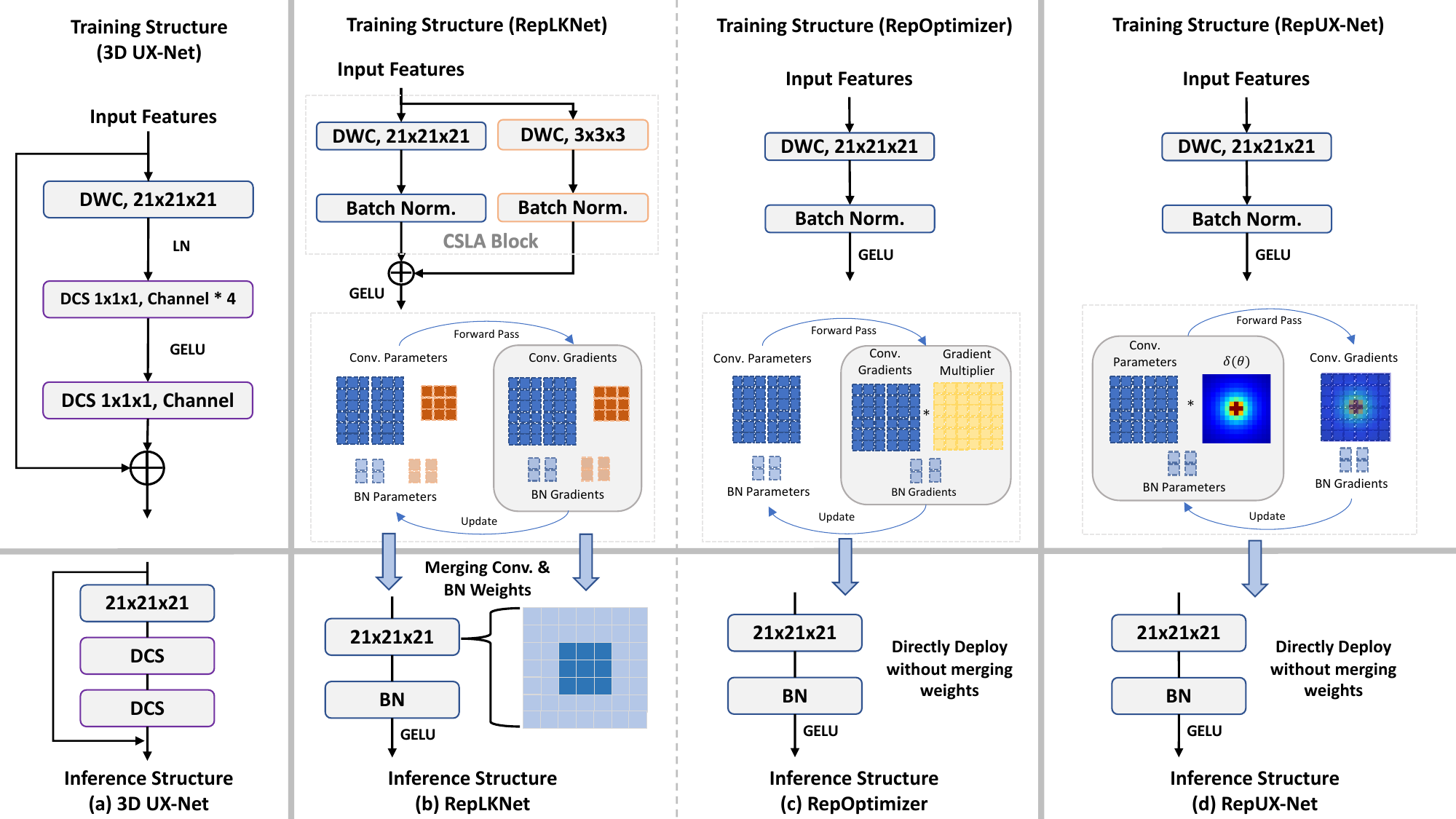}
\caption{Overview of RepUX-Net. Unlike performing SR to merge branches weight or performing GR within optimizers, we propose to multiply a Bayesian function $\delta$ and scale the element-wise learning importance in each large kernel. We then put the scaled weights back into the convolution layer for training.} \label{pipeline_figure}
\end{figure}

\section{Related Works}
\textbf{Weights Re-parameterization:} SR is a methodology of equivalently converting model structures via transforming the parameters in kernel weights. For example, RepVGG demonstrates to construct one extra ResNet-style shortcut as a $1\times 1$ convolution, parallel to $3\times 3$ convolution during training \cite{ding2021repvgg}. Such parallel branch design is claimed to enhance the learning efficiency during training, in which the 1x1 branch is then merged into the parallel $3\times 3$ kernel via a series of linear transformation in the inference stage. OREPA further adds more parallel branches with linear scaling modules to enhance training efficiency \cite{hu2022online}. Inspired by the parallel branches design, RepLKNet is proposed to scale up the 2D kernel size (e.g., 31x31) with a 3x3 convolution as the parallel branch \cite{ding2022scaling}. SLaK further extends the kernel size to 51x51 by decomposing the large kernel into two rectangular parallel kernels with sparse groups and training the model with dynamic sparsity \cite{liu2022more}. However, the proposed models’ FLOPs remain at a high-level with the parallel branch design and demonstrates to have a trade-off between model performance and training efficiency. To tackle the trade-off, RepOptimizer provides an alternative to re-parameterize the back-propagate gradient, instead of the structural parameters of kernel weights, to enhance the training efficiency with plain convolution block design \cite{ding2022re}. Significant efforts have been demonstrated to enlarge the 2D kernel size in the natural image domain, while limited studies have been proposed for 3D kernels in medical domain. As 3D kernels have a larger number of parameters than 2D, it is challenging to directly leverage the parallel branch design and maintain an optimal convergence of learning large kernel convolution without trading off the computation efficiency significantly. 

\section{Methods}
Instead of changing the gradient dynamics during training \cite{ding2022re}, we introduce RepUX-Net, a pure 3D CNN architecture that performs element-wise scaling in large kernel weights to enhance the learning convergence and effectively adapts large receptive field for volumetric segmentation. To design such behavior, we adapt a two-step pipeline: 1) we define the theoretical equivalency of variable learning convergence in convolution branches; 2) we simulate the behavior of spatial frequency to re-weight the learning importance of each element in kernels for stochastic gradient descent. Note the theoretical derivation depends on the optimization with first-order gradient-driven optimizer (e.g., SGD, AdamW) \cite{ding2022re}. 

\subsection{Variable Learning Convergence in Multi-Branch Design}
From Figure 2, the learning convergence of the large kernel convolution can be improved by either adding up the encoded outputs of parallel branches weighted by diverse scales with SR (RepLKNet \cite{ding2022scaling}) or performing Gradient Re-parameterization (GR) by multiplying with constant values (RepOptimizer \cite{ding2022re}) in a Single Operator (SO). Inspired by the concepts of SR and GR, we extend the equivalency proof in RepOptimizer to adapt variable learning convergence in branches. Here, we only showcase the conclusion with two convolutions and two constant scalars as the scaling factors for simplicity. The complete proof of equivalency is demonstrated in Supplementary 1.1. Let $\{\alpha_L, \alpha_S\}$ and $\{W_L, W_S\}$ be the two constant scalars and two convolution kernels (Large \& Small) respectively. Let $X$ and $Y$ be the input and output features, the CSLA block is formulated as $Y_{CSLA} = \alpha_L(X\star W_L) + \alpha_S(X\star W_S)$, where $\star$ denotes as convolution. For SO blocks, we train the plain structure parameterized by $W'$ and $Y_{SO} = X\star W'.$ Let $i$ be the number of training iterations, we ensure that ${Y^{(i)}}_{CSLA} = {Y^{(i)}}_{SO}, \forall i\geq 0$ and derive the stochastic gradient descent of parallel branches as follows:
% \begin{equation}
%     {\alpha_{L}}W_{L(i+1)} + {\alpha_{S}}W_{S(i+1)} = {\alpha_{L}}W_{L(i)} + {\alpha_{S}}W_{S(i)} -\lambda({\alpha_L}{\frac{\partial{\mathcal{L}}}{\partial{W_{L_i}}}}+{\alpha_{S}}\frac{\partial{\mathcal{L}}}{\partial{W_{S_i}}}),
% \end{equation}
% where $\mathcal{L}$ is the objective function and $\lambda$ as the learning rate. We further expand the bracket in eq. 1 and observe that the gradient optimization of each branch can be different by adjusting the branch-specific Learning Rate (LR) as follows:
\begin{equation}
    {\alpha_L}W_{L(i+1)} + {\alpha_S}W_{S(i+1)} = {\alpha_L}W_{L(i)} -{\lambda_L}{\alpha_L}{\frac{\partial{\mathcal{L}}}{\partial{W_{L_i}}}} + {\alpha_S}W_{S(i)} - {\lambda_S}{\alpha_S}{\frac{\partial{\mathcal{L}}}{\partial{W_{S_i}}}},
\end{equation}
where $\mathcal{L}$ is the objective function; $\lambda_L$ and $\lambda_S$ are the Learning Rate (LR) of each branch respectively. We observe that the optimization of each branch can be different by adjusting the branch-specific LR. The locality convergence in large kernels enhance with the quick convergence in small kernels. Additionally from our experiments, a significant improvement is demonstrated with different branch-wise LR using SGD (Table 2). With such observation, we further hypothesize that \textbf{the convergence of each large kernel element can be optimized differently by linear scaling with prior knowledge}.

\subsection{Bayesian Frequency Re-parameterization (BFR)}
With the visualization of ERF in RepLKNet \cite{ding2022scaling}, the diffused distribution (from local to global) in ERF demonstrates similar behavior with the spatial frequency in the human visual system \cite{kulikowski1982theory}. High spatial frequency (small ERF) allows to refine and sharpen details with high acuity, while global details are demonstrated with low spatial frequency. Inspired by the reciprocal characteristics in spatial frequency, we first generate a Bayesian prior distribution to model the spatial frequency by computing a reciprocal distance function between each element and the central point of the kernel weight as follows:
\begin{equation}
\begin{aligned}
    d(x, y, z, c) & = \sqrt {\left( {x- c} \right)^2 + \left( {y - c } \right)^2 + \left( {z - c } \right)^2} \\
    \delta(x_k, y_k, z_k, c, \alpha) & =  \frac{\alpha}{d(x_k, y_k, z_k, c) + \alpha}
\end{aligned}
\end{equation}
where $k$ and $c$ are the element and central index of the kernel weight, $\alpha$ is the hyperparameter to control the shape of the generated frequency distribution. Instead of adjusting the LR in parallel branches, we propose to re-parameterize the convolution weights by multiplying the scaling factor $\delta$ to each kernel element and apply a static LR $\lambda$ for stochastic gradient descent in single operator setting as follows:
\begin{equation}
    W^{'}_{i+1} = \delta W^{'}_i - \lambda{\frac{\partial{L}}{\partial{\delta W^{'}_i}}} 
\end{equation}
With the multiplication with $\delta$, each element in the kernel weight is rescaled with respect to the frequency level and allow to converge differently with a static LR in stochastic gradient descent. Such design demonstrates to influence the weighted convergence diffused from local to global in theory, thus tackling the limitation of enhancing the local convergence only in branch-wise setting.

\subsection{Model Architecture}
The backbone of RepUX-Net is based on 3D UX-Net \cite{lee20223d}, which comprises multiple volumetric convolution blocks that directly utilize 3D patches and leverage skip connections to transfer hierarchical multi-resolution features for end-to-end optimization. Inspired by \cite{li2022large}, we choose a kernel size of $21\times 21\times 21$ for DepthWise Convolution (DWC-21) as the optimal choice without significant trade-off between model performance and computational efficiency in 3D. We further simplify the block design as a plain convolution block design to minimize the computational burden from additional modules. The encoder blocks in layers $l$ and $l+1$ are defined as follows:
\begin{equation}
\begin{aligned}
   \hat{z}^{l}=\text{GeLU}(\text{DWC-21}(\text{BN}(z^{l-1}))),\: \hat{z}^{l+1}=\text{GeLU}(\text{DWC-21}(\text{BN}(z^{l})))
\end{aligned}
\end{equation}
where $\hat{z}_{l}$ and $\hat{z}_{l+1}$ are the outputs from the DWC layer in each depth level; BN denotes as the batch normalization layer.

\section{Experimental Setup}
\textbf{Datasets} We perform experiments on six public datasets for volumetric segmentation, which comprise with 1) Medical Segmentation Decathlon (MSD) spleen dataset \cite{antonelli2022medical}, 2) MICCAI 2017 LiTS Challenge dataset (LiTS) \cite{bilic2023liver}, 3) MICCAI 2019 KiTS Challenge dataset (KiTS) \cite{heller2020international}, 4) NIH TCIA Pancreas-CT dataset (TCIA) \cite{roth2015deeporgan}, 5) MICCAI 2021 FLARE Challenge dataset (FLARE) \cite{AbdomenCT-1K}, and 6) MICCAI 2022 AMOS challenge dataset (AMOS) \cite{ji2022amos}. More details of each dataset (including data split for training and inference) are described in Supplementary Material (SM) Table 1.\\
\textbf{Implementation} We evaluate RepUX-Net with three different scenarios: 1) internal validation with direct supervised learning, 2) external validation with the unseen datasets, and 3) transfer learning with pretrained weights. All preprocessing and training details including baselines, are followed with \cite{lee20223d} for benchmarking. For external validations, we leverage the AMOS-pretrained weights to evaluate 4 unseen datasets. In summary, we evaluate the segmentation performance of RepUX-Net by comparing current SOTA networks in a fully-supervised setting. Furthermore, we perform ablation studies to investigate the effect on Bayesian frequency distribution with different scales generated by $\alpha$ and the variability of branch-wise learning rates with first-order gradient optimizers (e.g., SGD, AdamW) for volumetric segmentation. Dice similarity coefficient is leveraged as an evaluation metric to measure the overlapping regions between the model predictions and the manual ground-truth labels. 

\begin{table*}[t!]
    \centering
    \caption{Comparison of SOTA approaches on the five different testing datasets. (*: $p< 0.01$, with Paired Wilcoxon signed-rank test to all baseline networks)}
    \begin{adjustbox}{width=\textwidth}
    \begin{tabular}{*{1}{l}|*{2}{c}|*{5}{c}|*{4}{c}}
        % \hline \hline
        \toprule
         & & & \multicolumn{5}{|c}{Internal Testing} & \multicolumn{4}{|c}{External Testing} \\
         & & & \multicolumn{5}{|c}{FLARE} & \multicolumn{1}{|c}{MSD} & KiTS & LiTS & TCIA \\
         \midrule
         Methods & \#Params & FLOPs & Spleen & Kidney & Liver & Pancreas & Mean & Spleen & Kidney & Liver & Pancreas\\
         \midrule
         % 3D U-Net \cite{cciccek20163d} & 4.81M & 135.9G & 0.911 & 0.962 & 0.905 & 0.789 & 0.892 \\
         % SegResNet \cite{myronenko20183d} & 1.18M & 15.6G & 0.963 & 0.934 & 0.965 & 0.745 & 0.902 \\
         % RAP-Net \cite{lee2021rap} & 38.2M & 101.2G & 0.946 & 0.967 & 0.940 & 0.799 & 0.913  \\
         nn-UNet \cite{isensee2021nnu} & 31.2M & 743.3G & 0.971 & 0.966 & 0.976 & 0.792 & 0.926 & 0.917 & 0.829 & 0.935 & 0.739\\
         \midrule
         TransBTS \cite{wang2021transbts} & 31.6M & 110.4G & 0.964 & 0.959 & 0.974 & 0.711 & 0.902 & 0.881 & 0.797 & 0.926 & 0.699\\
         UNETR \cite{hatamizadeh2022unetr} & 92.8M & 82.6G & 0.927 & 0.947 & 0.960 & 0.710 & 0.886 & 0.857 & 0.801 & 0.920 & 0.679\\
         nnFormer \cite{zhou2021nnformer} & 149.3M & 240.2G & 0.973 & 0.960 & 0.975 & 0.717 & 0.906 & 0.880 & 0.774 & 0.927 & 0.690\\
         SwinUNETR \cite{hatamizadeh2022swin} & 62.2M & 328.4G & 0.979 & 0.965 & 0.980 & 0.788 & 0.929 & 0.901 & 0.815 & 0.933 & 0.736 \\
         % UNesT (Large) & 279.5M & 598.0G & 0.881 & 0.766 & 0.932 & \textbf{0.874} & 0.915 & 0.885 & \textbf{0.847} & 0.871 \\
         3D UX-Net (k=7) \cite{lee20223d} & 53.0M & 639.4G & 0.981 & 0.969 & 0.982 & 0.801 & 0.934 & 0.926 & 0.836 & 0.939 & 0.750\\
         3D UX-Net (k=21) \cite{lee20223d} & 65.9M & 757.6G & 0.980 & 0.968 & 0.979 & 0.795 & 0.930 & 0.908 & 0.808 & 0.929 & 0.720 \\
         \midrule
         \textbf{RepOptimizer \cite{ding2022re}} & 65.8M & 757.4G & 0.981 & 0.969 & 0.981 & 0.822 & 0.937 & 0.913 & 0.833 & 0.934 & 0.746\\
         \textbf{3D RepUX-Net (Ours)} & 65.8M & 757.4G & \textbf{0.984} & \textbf{0.970} & \textbf{0.983} & \textbf{0.837} & \textbf{0.944*} & \textbf{0.932*} & \textbf{0.847*} & \textbf{0.949*} & \textbf{0.779*}\\
         \bottomrule
    \end{tabular}
    \end{adjustbox}
    \label{baselines_compare}
\end{table*}
\begin{table*}[t!]
    \centering
    \caption{Ablation studies with quantitative Comparison on Block Designs with/out frequency modeling using different optimizer}
    \begin{adjustbox}{width=0.85\textwidth}
    \begin{tabular}{*{1}{c}|*{3}{c}|*{3}{c}|*{1}{c}}
        % \hline \hline
        \toprule
        Optimizer\: & \:Main Branch & Para. Branch & BFR\: & \:Train Steps & Main LR & Para. LR\: & Mean Dice \\
        \midrule
        SGD & $21\times 21\times 21$ & $\times$ & $\times$ & 40000 & 0.0003 & $\times$ & 0.898\\
        AdamW & $21\times 21\times 21$ & $\times$ & $\times$ & 40000 & 0.0001 & $\times$ & 0.906\\
        SGD & $21\times 21\times 21$ & $3\times 3\times 3$ & $\times$ & 40000 & 0.0003 & 0.0006 & 0.917\\
        AdamW & $21\times 21\times 21$ & $3\times 3\times 3$ & $\times$ & 40000 & 0.0001 & 0.0001 & 0.929\\
        AdamW & $21\times 21\times 21$ & $\times$ & \checkmark & 40000 & 0.0001 & $\times$ & \textbf{0.938}\\
        \midrule
        SGD & $21\times 21\times 21$ & $3\times 3\times 3$ & $\times$ & 60000 & 0.0003 & 0.0006 & 0.930\\
        AdamW & $21\times 21\times 21$ & $3\times 3\times 3$ & $\times$ & 60000 & 0.0001 & 0.0001 & 0.938\\
        AdamW & $21\times 21\times 21$ & $\times$ & $\checkmark$ & 60000 & 0.0001 & $\times$ & \textbf{0.944}\\
        \bottomrule
    \end{tabular}
    \end{adjustbox}
    \label{nnUNet_compare}
\end{table*}

\section{Results}
\textbf{Different Scenarios Evaluations.} Table 1 shows the result comparison of current SOTA networks on medical image segmentation in a volumetric setting. With our designed convolutional blocks as the encoder backbone, RepUX-Net demonstrates the best performance across all segmentation task with significant improvement in Dice score (FLARE: 0.934 to 0.944, AMOS: 0.891 to 0.902). Furthermore, RepUX-Net demonstrates the best generalizability consistently with a significant boost in performance across 4 different external datasets (MSD: 0.926 to 0.932, KiTS: 0.836 to 0.847, LiTS: 0.939 to 0.949, TCIA: 0.750 to 0.779). Furthermore, from Figure 2A, RepUX-Net demonstrates the quickest convergence rate in training with AMOS datasets from scratch. For transfer learning scenario, the performance of RepUX-Net significantly outperforms the current SOTA networks with mean Dice of 0.911 (1.22\% enhancement), as shown in Table 2. RepUX-Net demonstrates its capabilities across the generalizability of unseen datasets and transfer learning ability. The qualitative representations (in SM Figure 1) further provides additional confidence of the quality improvement in segmentation predictions with RepUX-Net.\\
\textbf{Ablation studies with block designs \& optimizers.} With the plain convolution design, a mean dice score of 0.906 is demonstrated with AdamW optimizer and perform slightly better than that with SGD. With the additional design of a parallel small kernel branch, the segmentation performance significantly improved (SGD: 0.898 to 0.917, AdamW: 0.906 to 0.929) with the optimized parallel branch LR using SR. The performance is further enhanced (SGD: 0.917 to 0.930, AdamW: 0.929 to 0.937) without being saturated with the increase of the training steps. By adapting BFR, the segmentation performance outperforms the parallel branch design significantly with a Dice score of 0.944.\\
\textbf{Effectiveness on Different Frequency Distribution.} From Figure 2 in SM, RepUX-Net demonstrates the best performance when $\alpha=1$, while comparable performance is demonstrated in both $\alpha=0.5$ and $\alpha=8$. A possible family of Bayesian distributions (different shapes) may need to further optimize the learning convergence of kernels across each channel.\\
\textbf{Limitations.} The shape of the generated Bayesian distribution is fixed across all kernel weights with an unlearnable distance function. Each channel in kernels is expected to extract variable features with different distributions. Exploring different families of distributions to rescale the element-wise convergence in kernels will be our potential future direction.
\begin{table*}[t!]
\caption{Evaluations on the AMOS testing split in different scenarios.(*: $p< 0.01$, with Paired Wilcoxon signed-rank test to all baseline networks)}
\begin{adjustbox}{width=\textwidth}
\label{tab:btcv}
\begin{tabular}{l|ccccccccccccccc|c}
\toprule
\multicolumn{17}{c}{Train From Scratch Scenario} \\ 
\midrule
Methods & \multicolumn{1}{c}{Spleen} & \multicolumn{1}{c}{R. Kid} & \multicolumn{1}{c}{L. Kid} & \multicolumn{1}{c}{Gall.} & \multicolumn{1}{c}{Eso.} & \multicolumn{1}{c}{Liver} & \multicolumn{1}{c}{Stom.} & \multicolumn{1}{c}{Aorta} & \multicolumn{1}{c}{IVC} & \multicolumn{1}{c}{Panc.} & \multicolumn{1}{c}{RAG} & \multicolumn{1}{c}{LAG} & \multicolumn{1}{c}{Duo.} & \multicolumn{1}{c}{Blad.} & \multicolumn{1}{c|}{Pros.} & \multicolumn{1}{c}{Avg}\\ \midrule
\midrule
nn-UNet & 0.951 & 0.919 & 0.930 & 0.845 & 0.797 & 0.975 & 0.863 & 0.941 & 0.898 & 0.813 & 0.730 & 0.677 & 0.772 & 0.797 & 0.815 & 0.850 \\
\midrule
TransBTS & 0.930 & 0.921 & 0.909 & 0.798 & 0.722 & 0.966 & 0.801 & 0.900 & 0.820 & 0.702 & 0.641 & 0.550 & 0.684 & 0.730 & 0.679 & 0.783 \\
UNETR & 0.925 & 0.923 & 0.903 & 0.777 & 0.701 & 0.964 & 0.759 & 0.887 & 0.821 & 0.687 & 0.688 & 0.543 & 0.629 & 0.710 & 0.707 & 0.740 \\
nnFormer & 0.932 & 0.928 & 0.914 & 0.831 & 0.743 & 0.968 & 0.820 & 0.905 & 0.838 & 0.725 & 0.678 & 0.578 & 0.677 & 0.737 & 0.596 & 0.785   \\
SwinUNETR & 0.956 & 0.957 & 0.949 & 0.891 & 0.820 & 0.978 & 0.880 & 0.939 & 0.894 & 0.818 & 0.800 & 0.730 & 0.803 & 0.849 & 0.819 & 0.871 \\ 
3D UX-Net (k=7) & 0.966 & 0.959 & 0.951 & 0.903 & 0.833 & 0.980 & 0.910 & 0.950 & 0.913 & 0.830 & 0.805 & 0.756 & \textbf{0.846} & 0.897 & 0.863 & 0.890 \\ 
3D UX-Net (k=21) & 0.963 & 0.959 & 0.953 & \textbf{0.921} & 0.848 & 0.981 & 0.903 & 0.953 & 0.910 & 0.828 & 0.815 & 0.754 & 0.824 & 0.900 & 0.878 & 0.891 \\ 
RepOptimizer & 0.968 & \textbf{0.964} & 0.953 & 0.903 & 0.857 & 0.981 & 0.915 & 0.950 & 0.915 & 0.826 & 0.802 & 0.756 & 0.813 & 0.906 & 0.867 & 0.892 \\
\midrule
% RepUX-Net (RepOptimizer)& \textbf{0.972} & \textbf{0.963} & \textbf{0.964} & \textbf{0.911} & \textbf{0.861} & \textbf{0.982} & \textbf{0.921} & \textbf{0.956} & \textbf{0.924} & \textbf{0.837} & \textbf{0.818} & \textbf{0.777} & \textbf{0.831} & \textbf{0.916} & \textbf{0.879} & \textbf{0.901*} \\ 
RepUX-Net (Ours) & \textbf{0.972} & 0.963 & \textbf{0.964} & 0.911 & \textbf{0.861} & \textbf{0.982} & \textbf{0.921} & \textbf{0.956} & \textbf{0.924} & \textbf{0.837} & \textbf{0.818} & \textbf{0.777} & 0.831 & \textbf{0.916} & \textbf{0.879} & \textbf{0.902*} \\ 
\midrule
\multicolumn{17}{c}{Transfer Learning Scenario} \\ 
\midrule
Methods & \multicolumn{1}{c}{Spleen} & \multicolumn{1}{c}{R. Kid} & \multicolumn{1}{c}{L. Kid} & \multicolumn{1}{c}{Gall.} & \multicolumn{1}{c}{Eso.} & \multicolumn{1}{c}{Liver} & \multicolumn{1}{c}{Stom.} & \multicolumn{1}{c}{Aorta} & \multicolumn{1}{c}{IVC} & \multicolumn{1}{c}{Panc.} & \multicolumn{1}{c}{RAG} & \multicolumn{1}{c}{LAG} & \multicolumn{1}{c}{Duo.} & \multicolumn{1}{c}{Blad.} & \multicolumn{1}{c|}{Pros.} & \multicolumn{1}{c}{Avg}\\ \midrule
\midrule
nn-UNet & 0.965 & 0.959 & 0.951 & 0.889 & 0.820 & 0.980 & 0.890 & 0.948 & 0.901 & 0.821 & 0.785 & 0.739 & 0.806 & 0.869 & 0.839 & 0.878 \\
\midrule
TransBTS & 0.885 & 0.931 & 0.916 & 0.817 & 0.744 & 0.969 & 0.837 & 0.914 & 0.855 & 0.724 & 0.630 & 0.566 & 0.704 & 0.741 & 0.650 & 0.792 \\
UNETR & 0.926 & 0.936 & 0.918 & 0.785 & 0.702 & 0.969 & 0.788 & 0.893 & 0.828 & 0.732 & 0.717 & 0.554 & 0.658 & 0.683 & 0.722 & 0.762 \\
nnFormer & 0.935 & 0.904 & 0.887 & 0.836 & 0.712 & 0.964 & 0.798 & 0.901 & 0.821 & 0.734 & 0.665 & 0.587 & 0.641 & 0.744 & 0.714 & 0.790   \\
SwinUNETR & 0.959 & 0.960 & 0.949 & 0.894 & 0.827 & 0.979 & 0.899 & 0.944 & 0.899 & 0.828 & 0.791 & 0.745 & 0.817 & 0.875 & 0.841 & 0.880 \\ 
3D UX-Net (k=7) & 0.970 & 0.967 & 0.961 & 0.923 & 0.832 & 0.984 & 0.920 & 0.951 & 0.914 & 0.856 & 0.825 & 0.739 & 0.853 & 0.906 & 0.876 & 0.900 \\ 
3D UX-Net (k=21) & 0.969 & 0.965 & 0.962 & 0.910 & 0.824 & 0.982 & 0.918 & 0.949 & 0.915 & 0.850 & 0.823 & 0.740 & 0.843 & 0.905 & 0.877 & 0.898 \\ 
RepOptimizer & 0.967 & 0.967 & 0.957 & 0.908 & 0.847 & 0.983 & 0.913 & 0.945 & 0.914 & 0.838 & 0.825 & 0.780 & 0.836 & 0.915 & 0.864 & 0.897 \\
\midrule
RepUX-Net & \textbf{0.973} & \textbf{0.968} & \textbf{0.965} & \textbf{0.933} & \textbf{0.865} & \textbf{0.985} & \textbf{0.930} & \textbf{0.960} & \textbf{0.923} & \textbf{0.859} & \textbf{0.829} & \textbf{0.793} & \textbf{0.869} & \textbf{0.918} & \textbf{0.891} & \textbf{0.911*} \\ 

\bottomrule
\end{tabular}
\end{adjustbox}
\end{table*}

\section{Conclusion}
We introduce RepUX-Net, the first 3D CNN adapting extreme large kernel convolution in encoder network for medical image segmentation. We propose to model the spatial frequency in the human visual system as a reciprocal function, which generates a Bayesian prior to rescale the learning convergence of each element in kernel weights. By introducing the frequency-guided importance during training, RepUX-Net outperforms current SOTA networks on six challenging public datasets via both direct training and transfer learning scenarios.

\bibliographystyle{splncs04}
\bibliography{main_latest}

\newpage

\section{Supplementary Material}
\subsection{Derivation of Variable Convergence in Multi-Branch Design}
The parallel structural design is referred to the CSLA block. Each branch only comprises on differentiable linear operator with trainable parameters (e.g., Convolution (Conv), Fully-Connected (FC) layer, scaling layer) and no training-time non-linearity. We begin with a simple case where the CSLA block has two parallel Conv kernels with same dimensions in kernel weights by padding and scaled by constant values. Let {$\alpha_L, \alpha_S$} and $W_L$, $W_S$ be the constant scalars and the weights of two Conv kernels (Large \& Small), and $X$ and $Y$ be the input and the output features. The computation flow of the CSLA block is formulated as following:
\begin{equation}
    Y_{CSLA} = \alpha_L(X\ast W_L) + \alpha_S(X\ast W_S),
\end{equation}
where $\ast$ denotes convolution. To optimize the gradient in parallel structure as a Single Operator (SO), we first derive the gradient flow scenario in a single operator structure, which is parameterized by $W^{'}$ as follows:
\begin{equation}
    Y_{SO} = X\ast W^{'}
\end{equation}
Our goal is to ensure generating same outputs in both multi-branch and single operator setting $Y_{CSLA} = Y_{SO}$ during training. Since only linear operations are performed within branches, we derive the kernel weights relationship as follows:
\begin{equation}
    W^{'} = {\alpha_L}W_{L(i)} + {\alpha_S}W_{S(i)}.
\end{equation}
With the above theoretical relationship in kernel weights, we apply the stochastic gradient descent rule and update the parallel branches gradient as follows:
\begin{equation}
    W^{'}_{i+1} = W^{'}_i - \lambda{\frac{\partial{\mathcal{L}}}{\partial{W^{'}_i}}} 
\end{equation}
\begin{equation}
    {\alpha_L}W_{L(i+1)} + {\alpha_S}W_{S(i+1)} = {\alpha_L}W_{L(i)} + {\alpha_S}W_{S(i)} -\lambda({\alpha_L}{\frac{\partial{\mathcal{L}}}{\partial{W_{L_i}}}}+{\alpha_S}\frac{\partial{\mathcal{L}}}{\partial{W_{S_i}}})
\end{equation}
where $\mathcal{L}$ is the differentiable loss function, $i$ is the index number of training iterations and $\lambda$ is the Learning Rate (LR). We further expand equation 5 and observe that each conv branch can be optimized with different convergence rate by adjusting the corresponding LR as follows:
\begin{equation}
    {\alpha_L}W_{L(i+1)} + {\alpha_S}W_{S(i+1)} = {\alpha_L}W_{L(i)} -{\lambda_L}{\alpha_L}{\frac{\partial{\mathcal{L}}}{\partial{W_{L_i}}}} + {\alpha_S}W_{S(i)} - {\lambda_S}{\alpha_S}{\frac{\partial{\mathcal{L}}}{\partial{W_{S_i}}}},
\end{equation}
where $\lambda_L$ and $\lambda_S$ are the LR for the large kernel branch and small kernel branch respectively. After training, the small kernel parameters are merged onto the central point of the large kernels, which is equivalent to enhance the learning convergence in locality with single operator setting.

% \subsection{Public Datasets Details}
\begin{table*}[t!]
    \centering
    \caption{Complete overview of six public MICCAI challenge datasets}
    \begin{adjustbox}{width=\textwidth}
    \begin{tabular}{*{1}{l}|*{6}{c}}
        % \hline \hline
        \toprule
        Challenge & FLARE & AMOS & MSD & KiTS & LiTS & TCIA \\
        \midrule
        Imaging Modality & Multi-Contrast CT & Multi-Contrast CT & Venous CT & Arterial CT & Venous CT & Venous CT\\
        Anatomical Region & Abdomen & Abdomen & Spleen & Kidney & Liver & Pancreas\\
        Sample Size & 361 & 200 & 41 & 300 & 131 & 89 \\
        \midrule
        \multirow{3}{*}{Anatomical Label} & \multirow{3}{*}{Spleen, Kidney, Liver, Pancreas} & Spleen, Left \& Right Kidney, Gall Bladder, &  \multirow{3}{*}{Spleen} &  \multirow{3}{*}{Kidney, Tumor} &  \multirow{3}{*}{Liver, Tumor} &  \multirow{3}{*}{Pancreas} \\
        & & Esophagus, Liver, Stomach, Aorta, Inferior Vena Cava (IVC) & & & & \\
        & & Pancreas, Left \& Right Adrenal Gland (AG), Duodenum & & & & \\
        \midrule
        \multirow{2}{*}{Data Splits} & 5-Fold Cross-Validation (Internal) & 1-Fold (Internal) & \multicolumn{4}{c}{All (External)}\\
        & Train: 272 / Validation: 69 / Test: 20 & Train: 160 / Validation: 20 / Test: 20 & Test: 41 & Test: 300 & Test: 131 & Test: 89\\
        \bottomrule
    \end{tabular}
    \end{adjustbox}
    \label{baselines_compare}
\end{table*}

\begin{figure}[t!]
\centering
\includegraphics[width=\textwidth]{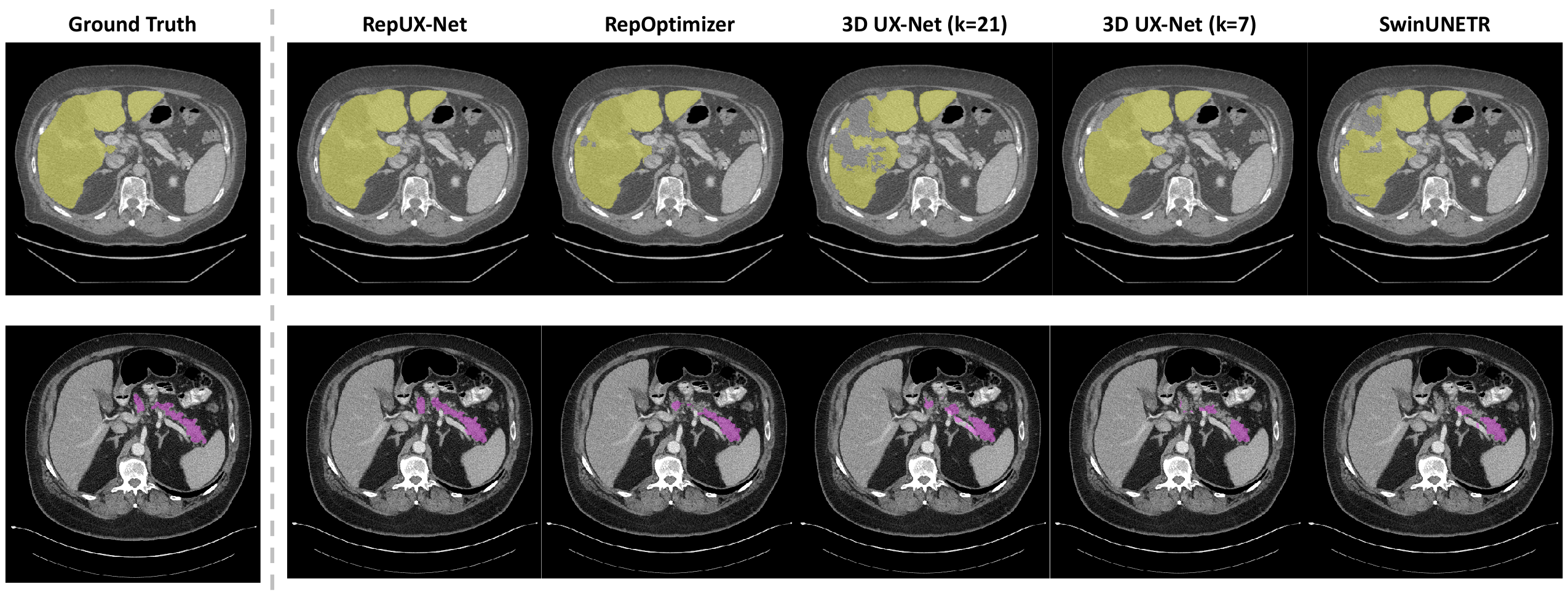}
\caption{Qualitative Representations of organ segmentation in LiTS and TCIA datasets} \label{quali_figure}
\end{figure}

\begin{figure}[t!]
\centering
\includegraphics[width=\textwidth]{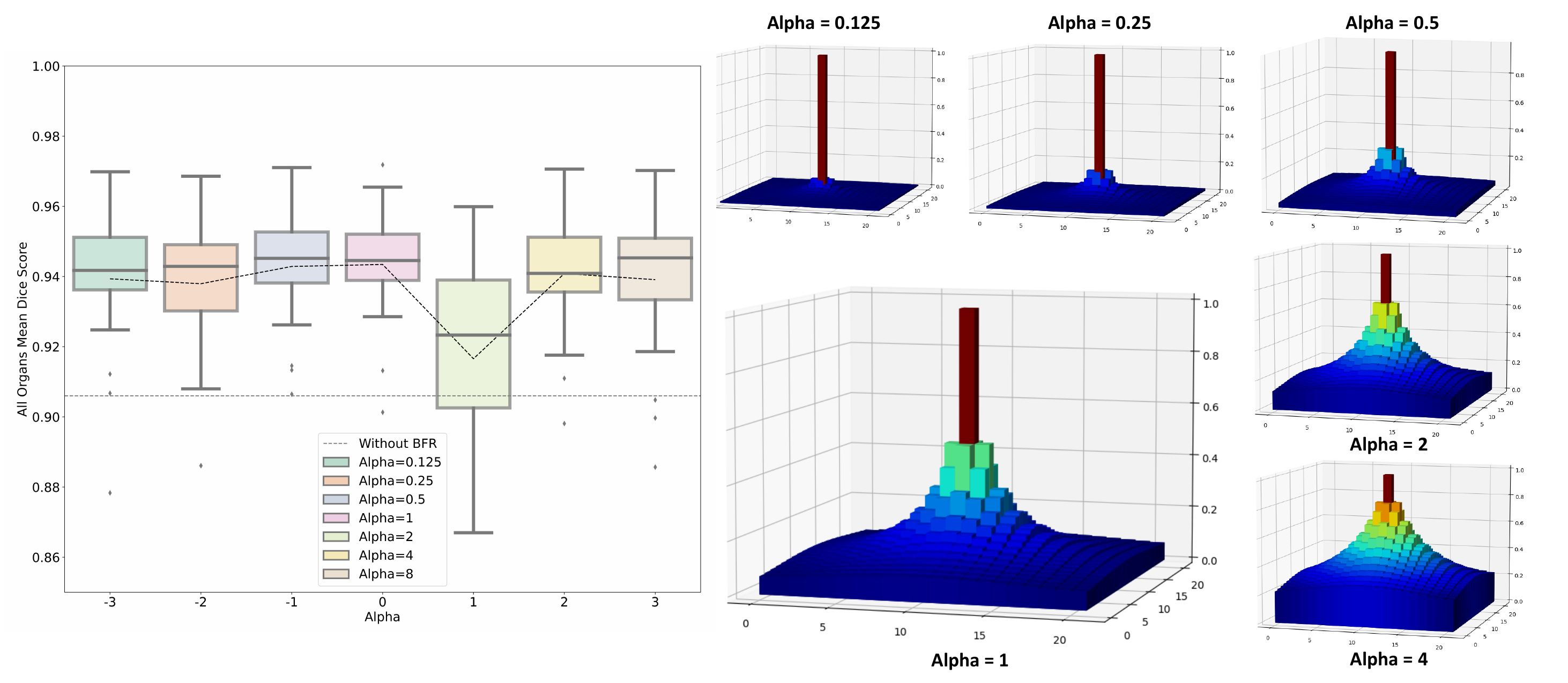}
\caption{Quantitative evaluation of the ablation study with different frequency distribution.} \label{quali_figure}
\end{figure}

\end{document}